%% file: EXO-24-004_temp.tex
\pdfoutput=1
\documentclass[11pt,twoside,a4paper,cmspaper,final,collab]{cms-tdr}
\def\svnVersion{563aa5e}\def\svnDate{2026/01/05}\def\cmsCernNoTag{CERN-EP-2025-166}\def\cmsCernDate{\today}\def\cmsMessage{Published in Physical Review D as \href{http://dx.doi.org/10.1103/7ldn-snzn}{\doi{10.1103/7ldn-snzn}.}}
\begin{document}\cmsNoteHeader{EXO-24-004}

\newlength\cmsTabSkip\setlength{\cmsTabSkip}{1ex}
\newcommand{\cbf}{\ensuremath{\mathcal{B}}\xspace}
\newcommand{\Nhits}{\ensuremath{N_\text{hits}}\xspace}
\newcommand{\vDPhi}{\ensuremath{\Delta\phi\text{(cluster, }\: \PGm_{\text{trigger}})}\xspace}
\newcommand{\IPSig}{\ensuremath{\mathrm{IP}_\text{sig}}\xspace}
\newcommand{\Zuu}{\ensuremath{\PZ\to\PGm\PGm}\xspace}
\newcommand{\BF}{\ensuremath{\cbf(\PB\to \PK \PGF )}\xspace}
\newcommand{\mP}{\ensuremath{m_{ \PGF }}\xspace}
\newcommand{\mk}{\ensuremath{m_{\PK}}\xspace}
\newcommand{\mb}{\ensuremath{m_{\PB}}\xspace}
\newcommand{\mz}{\ensuremath{m_{\PZ}}\xspace}
\newcommand{\ct}{\ensuremath{c\tau}\xspace}
\newcommand{\tcluster}{\ensuremath{t_\text{cluster}}\xspace}
\newcommand{\ifrac}[2]{\ensuremath{{#1}/{#2}}\xspace}

\newlength\cmsFigWidth
\ifthenelse{\boolean{cms@external}}{
  \setlength\cmsFigWidth{0.49\textwidth}
  \newcommand{\cmsLeft}{upper\xspace}
  \newcommand{\cmsRight}{lower\xspace}
 }{
  \setlength\cmsFigWidth{0.65\textwidth}
  \newcommand{\cmsLeft}{left\xspace}
  \newcommand{\cmsRight}{right\xspace}
 }

\cmsNoteHeader{EXO-24-004}

\title{Search for \texorpdfstring{\PQb}{b} hadron decays to long-lived particles in the CMS endcap muon detectors}

\date{\today}

\abstract{A search for long-lived particles originating from the decay of \PQb hadrons produced in proton-proton collisions with a center-of-mass energy of 13\TeV at the LHC is presented. The analysis is performed on a data set recorded in 2018, corresponding to an integrated luminosity of 41.6\fbinv. Interactions of the long-lived particles in the CMS endcap muon system would create hadronic or electromagnetic showers, producing clusters of detector hits. Selected events contain at least one such high-multiplicity cluster in the muon endcaps and require the presence of a displaced muon. The most stringent upper limits to date on the branching fraction \BF, where the long-lived particle \PGF decays to a pair of hadrons, are obtained for \PGF masses of 0.3--3.0\GeV and \PGF mean proper decay lengths in the range of 1--500\unit{cm}.
}

\hypersetup{
  pdfauthor={CMS Collaboration},
  pdftitle={Search for b hadron decays to long-lived particles in the CMS endcap muon detectors},
  pdfsubject={CMS},
  pdfkeywords={CMS, long-lived particle, B meson, muon system}
}

\maketitle

\section{Introduction}
Long-lived particles (LLPs) are predicted by many theoretical models that address shortcomings of the standard model (SM) of particle physics.
These include supersymmetric (SUSY) models such as mini-split  SUSY~\cite{Arkani-Hamed:2012fhg, Arvanitaki:2012ps}, gauge-mediated SUSY breaking~\cite{Giudice:1998bp}, $R$-parity-violating SUSY~\cite{Barbier:2004ez,Csaki:2013jza}, and stealth SUSY~\cite{Fan:2011yu,Fan:2012jf};
models addressing the hierarchy problem such as neutral naturalness~\cite{Chacko:2005pe, Burdman:2006tz,Cai:2008au,Chacko:2015fbc} and hidden valleys~\cite{Strassler:2006im,Strassler:2006ri};
models addressing dark matter~\cite{Baumgart:2009tn,Kaplan:2009ag,Chan:2011aa,Dienes:2011ja,Dienes:2012yz} and the matter-antimatter asymmetry of the Universe~\cite{Cui:2014twa};
and models that generate neutrino masses~\cite{Helo:2013esa,Batell:2016zod}.
In particular, so-called minimal versions of the aforementioned models, which extend the SM via a single new particle that may be a scalar, pseudoscalar, fermion, or vector state, are receiving increasing interest~\cite{Aielli:2019ivi}.

A common feature of the minimal models that introduce a new scalar particle (\PGF) is that this new particle mixes kinetically with the SM Higgs boson.
Through this mixing, the production cross sections and decay rates of the new scalar particle mirror those of the Higgs boson at a different mass.
If \PGF has a mass smaller than that of a \PB meson, it will be copiously produced in \PB meson decays~\cite{Grinstein:1988yu}.
The new scalar particle is also neutral under SM gauge interactions and can have a measurable lifetime~\cite{Bezrukov:2009yw}.
The sensitivity of the CERN LHC experiments to such models has previously been studied~\cite{Winkler:2018qyg,Fradette:2017sdd}.
Searches for long-lived scalar particles produced in the decay of a \PB meson have been performed by the CHARM~\cite{CHARM:1985anb}, Belle~\cite{Belle:2010ryo}, BaBar~\cite{BaBar:2015jvu}, and LHCb~\cite{LHCb:2016awg,LHCb:2015nkv} Collaborations, primarily targeting final states where the scalar particle decays into a pair of leptons.
In particular, the LHCb search in the dimuon final state provides the most stringent overall bounds.

\begin{figure}[htpb]
  \centering
  \includegraphics[width=0.45\columnwidth]{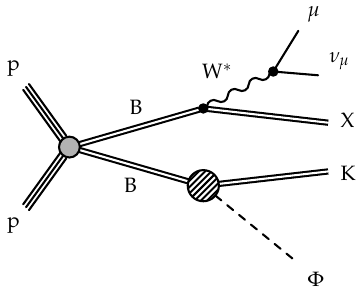}\hfill
  \includegraphics[width=0.45\columnwidth]{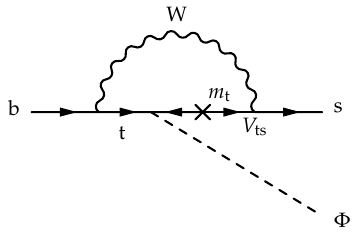}
  \caption{Example Feynman diagram for the production of a pair of \PB mesons, where one \PB meson decays to a muon, neutrino, and another hadron labeled \PX, while the other \PB meson decays to a kaon and the LLP \PGF (left).
    Penguin diagram displaying the flavor-changing interaction which changes the flavor of the \PQb quark to a \PQs quark, producing the  \PGF particle (right). The interaction proceeds via a virtual top quark, denoted by the cross, which undergoes a chirality flip, denoted by the outward-pointing arrows on either side of the cross in the diagram.}
  \label{fig:Feyn}
\end{figure}

Figure~\ref{fig:Feyn} (left) illustrates \PGF production through \PB meson pair production at the LHC.
One of the \PB mesons decays to the muon that will be used for the initial event selection, while the other \PB meson produces the \PGF.
The \PQb quark in this \PB meson undergoes a flavor-changing interaction, as illustrated in Fig.~\ref{fig:Feyn} (right).
A virtual top quark provides the coupling to the scalar LLP \PGF while the \PQb quark is transformed into an \PQs quark, which forms the kaon in Fig.~\ref{fig:Feyn} (left).
In the scenario where $\mk  \lesssim \mP  \lesssim  \mb$, where \mk (\mb) is the mass of the kaon (\PB meson), this \PGF production mode dominates~\cite{Willey:1982mc,Grinstein:1988yu}.
The LLP decays predominantly into a pair of pions or kaons, which, depending on the LLP's lifetime, may induce particle showers in different detector systems at the LHC experiments.
Decays to charged pions and kaons will induce hadronic showers while decays to neutral pions, which subsequently decay to a pair of photons, will induce electromagnetic showers. 
Although the description of the LLP production process above focuses on \PB meson production, the LLP and the shower signature may be created by any \PQb hadron generically, so the rest of this paper will use the term \PQb hadron and the symbol \PB to refer to the particle producing the LLP regardless of the \PQb hadron species.

In this paper, we present a search for LLPs produced in \PQb hadron decays that generate hadronic and electromagnetic showers in the endcap muon detectors of the CMS experiment.
This search uses the unique \PB parking data set~\cite{CMS:2024zhe} of proton-proton ($\Pp\Pp$) collisions at $\sqrt{s}=13\TeV$, which was collected in 2018 and corresponds to an integrated luminosity of 41.6\fbinv.

The search uses the muon detector shower (MDS) signature, defined as a cluster of reconstructed hits in the CMS muon system, which has previously been used by CMS to search for pair-produced LLPs decaying to hadrons~\cite{CMS:2021juv,CMS:2024bvl,CMS:2024ake}.
Because of the large amount of shielding provided by the steel return yoke of the CMS magnet, this signature yields very low levels of background from SM processes occurring within the bulk of the detector.
Showers are induced by the interaction of decay products with the steel absorbing layers, however these showers will be suppressed when the steel layer is very thick.
Such cases, in which a smaller fraction of the energy is deposited in the muon system, are denoted as ones with a diminished or low energy response. 
This is especially relevant for electromagnetic showers involving photons and electrons.

In this paper, Section~\ref{sec:cms} gives a brief description of the CMS detector. 
Section~\ref{sec:data} provides an overview of the data sets and simulated samples used to obtain the paper result. 
Section~\ref{sec:reco} describes how collision events are  reconstructed and how the MDS cluster signature, jets, muons, and other particle candidates and physical quantities are defined. 
The description of the strategy used for this search as well as the definition of the physics objects selections are provided in Section~\ref{sec:strategy}.
The estimation of the background is described in Section~\ref{sec:bkg}, and the treatment of systematic uncertainties is given in Section~\ref{sec:unc}. 
Results of this search and a summary of the paper are provided in Section~\ref{sec:results} and Section~\ref{sec:summary}, respectively.

The tabulated results for this paper along with signal efficiencies for the signal models considered can be found in the associated HEPData record~\cite{hepdata}.
\section{The CMS detector}
\label{sec:cms}

The CMS apparatus~\cite{CMS:2008xjf,CMS:2023gfb} is a multipurpose, nearly hermetic detector, designed to trigger on~\cite{CMS:2016ngn,CMS:2020cmk,CMS:2024aqx} and identify electrons, muons, photons, and (charged and neutral) hadrons~\cite{CMS:2014pgm,CMS:2018rym,CMS:2020uim}.
Its central feature is a superconducting solenoid of 6\unit{m} internal diameter, providing a magnetic field of 3.8\unit{T}.
Within the solenoid volume are a silicon pixel and strip tracker, a lead tungstate crystal electromagnetic calorimeter, and a brass and scintillator hadron calorimeter, each composed of a barrel and two endcap sections. Forward calorimeters extend the pseudorapidity ($\eta$) coverage provided by the barrel and endcap detectors.
Muons are reconstructed using gas-ionization detectors embedded in the steel flux-return yoke outside the solenoid using three technologies: drift tubes (DTs) in the barrel, cathode strip chambers (CSCs) in the
endcaps, and resistive-plate chambers (RPCs) in the barrel and endcaps.
More detailed descriptions of the CMS detector, together with a definition of the coordinate system used and the relevant kinematic variables, can be found in Refs.~\cite{CMS:2008xjf,CMS:2023gfb}.

The CMS muon detector system is an essential component of this search.
In the barrel region, the DTs are organized into four concentric cylindrical layers, called stations, separated by layers of the steel magnetic return yoke.
Starting from the innermost, the muon barrel stations are labeled MB1, MB2, MB3, and MB4.
The DTs cover the region $\abs{\eta} < 1.2$.
The DTs are chambers filled with gas, each with an anode wire at its center and cathodes arranged on the walls.
When a charged particle enters the tube, it ionizes the gas, liberating electrons that drift towards the anode at a constant velocity known as the drift velocity.
The position of the particle is obtained in the $\mathrm{R}$-$\phi$ and $\mathrm{R}$-$\theta$ planes (where R is radial, $\phi$ is azimuthal angle in radians, and $\theta$ is polar angle) using the known position of the individual DT cells and the distance from the anode to the point at which the particle enters the cell.
The CSCs are installed in the muon endcap and are arranged into four stations labeled ME1, ME2, ME3, and ME4.
The CSCs are positioned in a series of concentric rings around the beamline at each station.
Each station is separated into two rings in the radial direction except for the first station, which has three rings.
These ring subsections are labeled ME$N$/$n$ for ring $N$ and subsection $n$.
The CSCs have alternating layers of anode wires and cathode strips, arranged orthogonally. 
When a charged particle ionizes the gas, electrons move toward the wires while their motion induces image charges on the strips, providing two positional coordinates, in the  $\mathrm{R}$-$\phi$ plane, for the charged particle while the final coordinate along the $z$-axis is given by the location of the CSC chamber itself.
When combining the information from the signals in both the anode wires and the cathode strips, the spatial and time coordinates of the hit can be determined with the resolutions of 400-500\mum and 5\unit{ns} respectively~\cite{CMS:2018rym,CMS:2009bjz}. 
The CSCs cover the range $0.9 < \abs{\eta} < 2.4$.
Additional information on the DTs and CSCs can be found in Ref.~\cite{CMS:2021yvr}.

The RPCs are constructed with two parallel plates of highly resistive Bakelite, coated with a conductive paint to provide an anode and a cathode.
When a charged particle interacts with the gas in the chamber, it ionizes the gas triggering an avalanche of electrons.
The resulting charge buildup is collected at the anode and cathode plates.
The timing resolution of the RPCs is approximately 3\unit{ns}, allowing for the unambiguous assignment of hits to the appropriate bunch crossing~\cite{CMS:1997dma}.
The RPCs are attached to one or more sides of the CSC and DT layers and cover the range of $\abs{\eta} < 1.9$.
The RPCs in the endcap are labeled RE$N$/$n$ for ring $N$ and subsection $n$, similar to the CSCs.
In most stations, the RPCs are installed just behind the corresponding CSC station except for the RPCs in ring 2, which are installed just in front of the CSCs in that ring.
In the barrel region, RPCs are installed in front and behind the DT stations, except for the outermost layer where RPCs are only located in  front of the DT stations, and are labeled RB$M$ where $M$ is the station number from 1 to 4. 
Additional details on the RPCs can be found in Refs.~\cite{CMS:2008xjf, CMS:1997dma}. 

Events of interest are selected using a two-tiered trigger system.
The first level, composed of custom hardware processors, uses information from the calorimeters and muon detectors to select events at a rate of around 100\unit{kHz} within a fixed latency of about 4\mus~\cite{CMS:2020cmk}.
The second level, known as the high-level trigger, consists of a farm of processors running a version of the full event reconstruction software optimized for fast processing, and reduces the event rate to a few kHz before data storage~\cite{CMS:2016ngn,CMS:2024aqx}.

\section{Data and simulated samples}
\label{sec:data}

A novel method has been devised to record and process the data with the purpose of collecting a large sample of \bbbar events.
The \PB parking data set was collected using
a set of high-rate triggers that require the presence of a single displaced muon.
Here \textit{displaced} refers to the measure of the significance of the transverse impact parameter \IPSig of the track used to reconstruct the muon.
Depending on the running conditions, the minimum thresholds used on the transverse momentum (\pt) of the muon ranged from 7 to 12\GeV and on \IPSig ranged  from 3 to 6.
Such nonzero \IPSig values are characteristic of displaced muons from $\PQb\to \PGm\PX$ decays because of the relatively long lifetime of $\PQb$ hadrons.
Since \PQb hadrons are predominantly produced in pairs at the LHC, applying the trigger requirements to the decay of one of the \PQb hadrons allows us to save a large number of unbiased decays of the other \PQb hadron.
The result is a highly pure sample of $\sim$$10 \times 10^{9}$ unbiased \PQb hadron decays.
The resulting purity---the ratio of the number of \bbbar pairs to the number of $\Pp\Pp$ collisions recorded by the trigger system---in this data sample is measured in simulation and ranges from 59 to 92\% depending on the choice of muon parameters used in the trigger definition~\cite{CMS:2024zhe}.

The \PB parking data set, though originally produced to study lepton flavor universality, provides a unique opportunity to search for LLPs.
In this investigation the LLP is assumed to be uncharged.
Since a neutral LLP does not interact directly in the detector, one must develop a trigger strategy that exploits other indirect features to select events containing the LLP and its decay.
Hence, without the presence of a dedicated trigger, searches for LLPs in CMS have relied on the use of triggers requiring large missing transverse momentum (\ptmiss) or associated production with another particle, such as a high-\pt lepton.
This indirect trigger strategy generally leads to lower selection efficiencies for LLPs compared to the \PB parking strategy used in this work.
For example, the \PB parking trigger strategy provides approximately a 10\% higher signal efficiency than a trigger strategy based on \ptmiss for the model and signature described in this paper.

Several simulated samples are produced to model the $\PB\to \PK\PGF$ process.
Samples are generated for \PB meson pair production, where $\PB \to \PGm \PGnGm \PX$ for one of the \PB mesons and $\PB\to \PK\PGF$ for the other, assuming five different \PGF mass values, $\mP = 0.3$, 0.5, 1.0, 2.0 and 3.0\GeV.
The samples are generated inclusively, containing decays of $\PBp$, $\PBm$, and neutral \PB mesons, in proportion to cross sections derived independently from the \PB parking data.
In simulation, the \PGF particle is modeled to decay exclusively via hadronic modes, producing either quark-antiquark pairs, or mesons or baryons directly.
For each sample, the mean proper decay length of \PGF varies in the range $1 < \ct < 500\unit{cm}$.
For each unique \mP and decay mode (hadronic or electromagnetic shower), samples are generated at two values of \ct: one chosen to yield the highest expected signal acceptance in the muon system for the given \mP, and another, an order of magnitude larger.
Additional signal hypotheses at other \ct values are derived by reweighting the distribution of the proper decay length.
The $\Pp\Pp$ collision and hadronization process is generated with \PYTHIA8.230~\cite{Sjostrand:2014zea}, using the underlying event tune
CP5~\cite{CMS:2019csb}, and the NNPDF3.1 parton distribution functions at next-to-next-to-leading order with the strong coupling constant $\alpS(\mz) = 0.118$, where \mz denotes the \PZ boson mass~\cite{NNPDF:2017mvq}.
The decays of \PB mesons are generated using \EVTGEN1.30~\cite{Lange:2001uf}.
The detector response is simulated using a detailed description of the CMS detector based on \GEANTfour~\cite{GEANT4:2002zbu}.
In our simulation, the \PGF particle is implemented as a stable, non-interacting neutral state until it decays, ensuring that only its decay products interact with the detector.
The branching fractions for the decay of the light scalar particle \PGF were determined following the work in Refs.~\cite{Winkler:2018qyg,Gershtein:2020mwi}.
Further details can be found in Ref.~\cite{Mitridate:2023tbj}.

\section{Event reconstruction}
\label{sec:reco}
The global event reconstruction (also called particle-flow event reconstruction~\cite{CMS:2017yfk}) aims to reconstruct and identify each individual particle in an event, with an optimized combination of all subdetector information.
In this process, the identification of the particle type (photon, electron, muon, charged or neutral hadron) plays an important role in the determination of the particle direction and energy.
The primary vertex is taken to be the vertex corresponding to the hardest scattering in the event, evaluated using tracking information alone, as described in Section 9.4.1 of Ref.~\cite{CMS-TDR-15-02}.

Muons (\eg, from \PQb hadron semileptonic decays) are identified as tracks in the central tracker consistent with either a track or several hits in the muon system, and associated with calorimeter deposits compatible with the muon hypothesis.
The energy of muons is obtained from the corresponding track momentum.
Muons are measured in the range $\abs{\eta} < 2.4$.
The single-muon trigger efficiency exceeds 90\% over the full $\eta$ range, and the efficiency to reconstruct and identify muons is greater than 96\%.
Matching muons to tracks measured in the silicon tracker results in a relative \pt resolution of 1\% in the barrel and 3\% in the endcaps for muons with $\pt< 100$\GeV~\cite{CMS:2018rym}.

For each event, hadronic jets are clustered from the reconstructed particles using the infrared- and collinear-safe anti-\kt algorithm~\cite{Cacciari:2008gp, Cacciari:2011ma} with a distance parameter of 0.4.
Jet momentum is determined as the vectorial sum of all particle momenta in the jet, and is found from simulation to be, on average, within 5--10\% of the true momentum over the whole \pt spectrum and detector acceptance.
Additional $\Pp\Pp$ interactions within the same or nearby bunch crossings are known as pileup interactions.
Pileup can contribute additional tracks and calorimetric energy depositions to the jet momentum.
To mitigate this effect, charged particles identified to be originating from pileup vertices are discarded and an offset correction is applied to account for remaining contributions~\cite{Sirunyan:2020foa}.
Jet energy corrections are derived from simulation to bring the measured response of jets to that of particle level jets on average.
In situ measurements of the momentum balance in dijet, photon+jet, $\PZ$+jet, and multijet events are used to account for any residual differences in the jet energy scale between data and simulation~\cite{CMS:2016lmd}.
The jet energy resolution amounts typically to 15--20\% at 30\GeV and 10\% at 100\GeV~\cite{CMS:2016lmd}.
Additional selection criteria are applied to each jet to remove jets potentially dominated by anomalous contributions from various subdetector components or reconstruction failures~\cite{CMS:2011bgj}.
Hadronic jets can produce particle showers in the muon detectors similar to those of signal processes and must be accounted for separately, as discussed in Section~\ref{sec:bkg}.

The vector \ptvecmiss is computed as the negative vector \pt sum of all the reconstructed particles in an event~\cite{CMS:2019ctu}.
The \ptvecmiss is modified to account for corrections to the energy scale of the reconstructed jets in the event.

\begin{figure*}
  \centering  
  \includegraphics[width=0.9\textwidth]{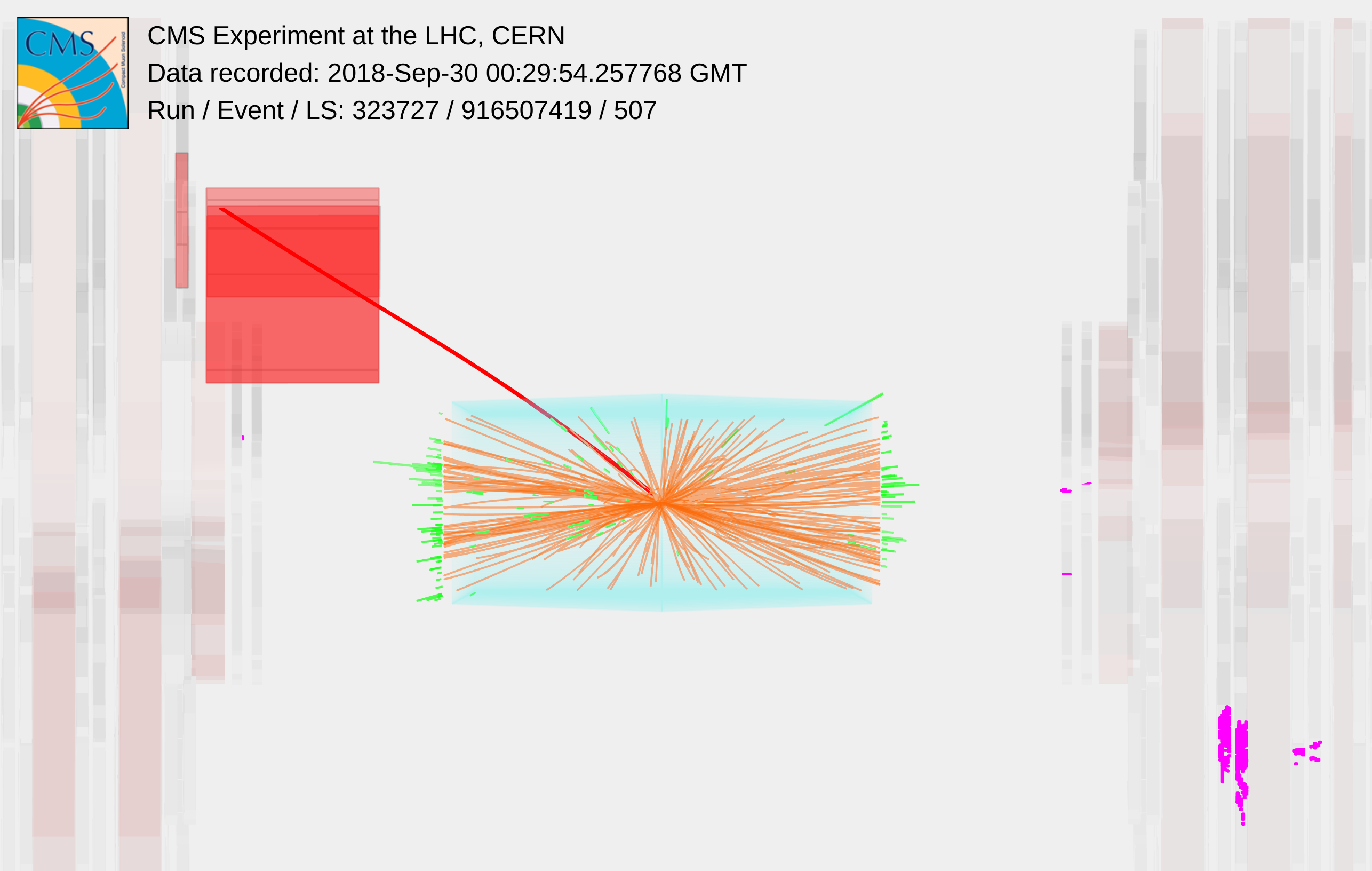}
    \caption{Display of a candidate signal event in the \PB parking data set.
    The display shows the hits (magenta points) in the muon endcap system that form the MDS cluster opposite the muon (red line), which is used to select the event on-line.
    Muon system chambers that recorded hits from a muon are shown as red rectangles. }
    \label{fig:display}
\end{figure*}

The hadronic and electromagnetic decays of hadrons produced by the \PGF particle provide the MDS signature used in this analysis.
Clusters are composite objects constructed from the reconstructed hits in the CSCs.
Clusters may be formed within a single CSC module, or span multiple modules located at different positions along the beam axis but occupying the same $\eta$--$\phi$ region.
Clustering of the hits in the muon system is achieved using the density-based spatial clustering of applications with noise (dbscan) algorithm~\cite{dbscan}.
The clustering parameter $\DR=\sqrt{\smash[b]{(\Delta\eta)^2 +(\Delta\phi)^2}}$ is required to be less than 0.2 to group hits or groups of hits into clusters.
Cluster-level quantities, such as the $\eta$ and $\phi$ coordinates of a cluster, are computed based on the average position of the hits in Cartesian coordinates used to construct the cluster object.
To suppress the background from muons traversing the detector, we apply a minimum threshold of 50 hits within a candidate cluster because a minimum ionizing muon is expected to produce approximately 24 hits in the CSCs~\cite{CMS:2021juv,CMS:2024bvl}.
Figure~\ref{fig:display} depicts a candidate signal event from the \PB parking data set.
The muon used to select the event on-line is shown with a red line.
Opposite the muon is a set of hits, illustrated as magenta points, in the CSCs spanning multiple stations of the endcap detector, forming the MDS cluster. 

\section{Search strategy and selections}
\label{sec:strategy}

There are two main classes of background for this analysis.
The first, the uncorrelated background, originates from prompt particles that produce a shower in the muon system.
This background is composed of events containing muon bremsstrahlung radiation, particles originating from adjacent bunch crossings and particles originating from secondary interactions within the detector.
This background is largely mitigated through the use of physics object selections and vetoes.
The other source of background, the jet-induced background, is due to long-lived SM particles decaying to jets, including SM decays of the \PQb hadron, such as $\PB\to \PGm \PGnGm \PX$ and $\PB\to \PK\PX$.
These jets are referred to as punch-through jets---jets produced in the bulk of the CMS detector, but not stopped by the outer steel layer.
In this scenario, these jets may produce a signal-like cluster in the muon system.
This jet-induced background is irreducible because it shares features with the signal process.
To account for the jet-induced background, a ``misidentification rate'' estimate was developed to measure its contribution in the signal region (SR).

Events are selected based on the presence of a displaced muon, as described in Section~\ref{sec:data}.
Muon candidates in this analysis are required to have $\pt > 7\GeV$ and $\abs{\eta} < 1.5$ to satisfy the trigger requirements.
Additionally, the track associated with the muon candidate must contain at least five hits in the strip tracking layer and at least one hit within the pixel tracking layer and the track must be matched to at least one segment in the muon system.
A segment is a localized muon track candidate reconstructed from hits within the muon system that correlate with the trajectory of a muon~\cite{CMS:2006myw}.
The muon candidate must also have a longitudinal impact parameter $\abs{d_z} < 20\unit{cm}$ and a transverse impact parameter $\abs{d_{xy}} < 0.3\unit{cm}$ as measured from the primary vertex.
These requirements are chosen to be consistent with the production of muons from the decays of \PQb hadrons~\cite{CMS:2017yfk}.
Events are also required to have a cluster in the CSCs.
A number of geometric selections are applied to reject clusters created by punch-through jets.
This is most prevalent in the first layers of the muon system, just after the first steel layer.
Clusters are vetoed if they contain any hits in either the ME1/1 or ME1/2 rings of the CSC system (ME11/12 veto).
An additional veto is applied on clusters matched to RE1/2 chambers.
This selection will be referred to as the RE1 veto.
The RE1 veto is applied to remove punch-through jets causing clusters in the CSCs that are not suppressed by the ME11/12 veto.
In addition, to suppress clusters caused by muon bremsstrahlung in the region $0.8 < \abs{\eta} < 1.2$, where the muons traverse the barrel DT stations and shower in ME1/3 or ME2/2, we veto clusters that are matched to DT segments in the first DT station (MB1) within $\DR < 0.4$ or are matched to the RPC hits in the first barrel RPC station (RB1) within $\DR < 0.4$.
These are respectively referred to as the MB1 segment and RB1
vetoes.
We veto clusters with $\abs{\eta} > 1.9$ to account for cases where a muon is not reconstructed successfully.
In this case, a shower produced by the misreconstructed muon may appear as an MDS and would not be rejected by other vetoes.
Finally, energetic muons may undergo the bremsstrahlung process and create electromagnetic showers in the CSCs.
To mitigate this muon bremsstrahlung background, we veto clusters matched to a muon with $\pt> 20\GeV$ within $\DR < 0.5$ (muon bremsstrahlung veto).

We also apply a cluster identification algorithm for clusters in the region of $\abs{\eta} < 1.9$. 
In earlier results using the MDS signature, it was found that clusters from the uncorrelated background occur more frequently at large values of $\abs{\eta}$, where reconstruction inefficiencies become more prevalent, reducing the effectiveness of the muon bremsstrahlung veto~\cite{CMS:2021juv,CMS:2024bvl,CMS:2024ake}. 
This algorithm places stricter $\abs{\eta}$ requirements on the cluster when it is closer to the primary interaction point or if the cluster occupies only one station ($N_\text{stations} = 1$).
The requirements on $\abs{\eta}$ are as follows:
\begin{itemize}
    \item $\abs{\eta} < 1.9$ if $N_\text{stations} > 1$,
    \item $\abs{\eta} < 1.8$ if $N_\text{stations} = 1$ and the cluster is in station 4,
    \item $\abs{\eta} < 1.6$ if $N_\text{stations} = 1$ and the cluster is in station 3,
    \item $\abs{\eta} < 1.7$ if $N_\text{stations} = 1$ and the cluster is in station 2,
    \item $\abs{\eta} < 1.1$ if $N_\text{stations} = 1$ and the cluster is in station 1 because of an implicit selection from the ME1/1 or ME1/2 vetoes.
\end{itemize}
\begin{figure}[!htb]
  \centering
    \includegraphics[width=\cmsFigWidth]{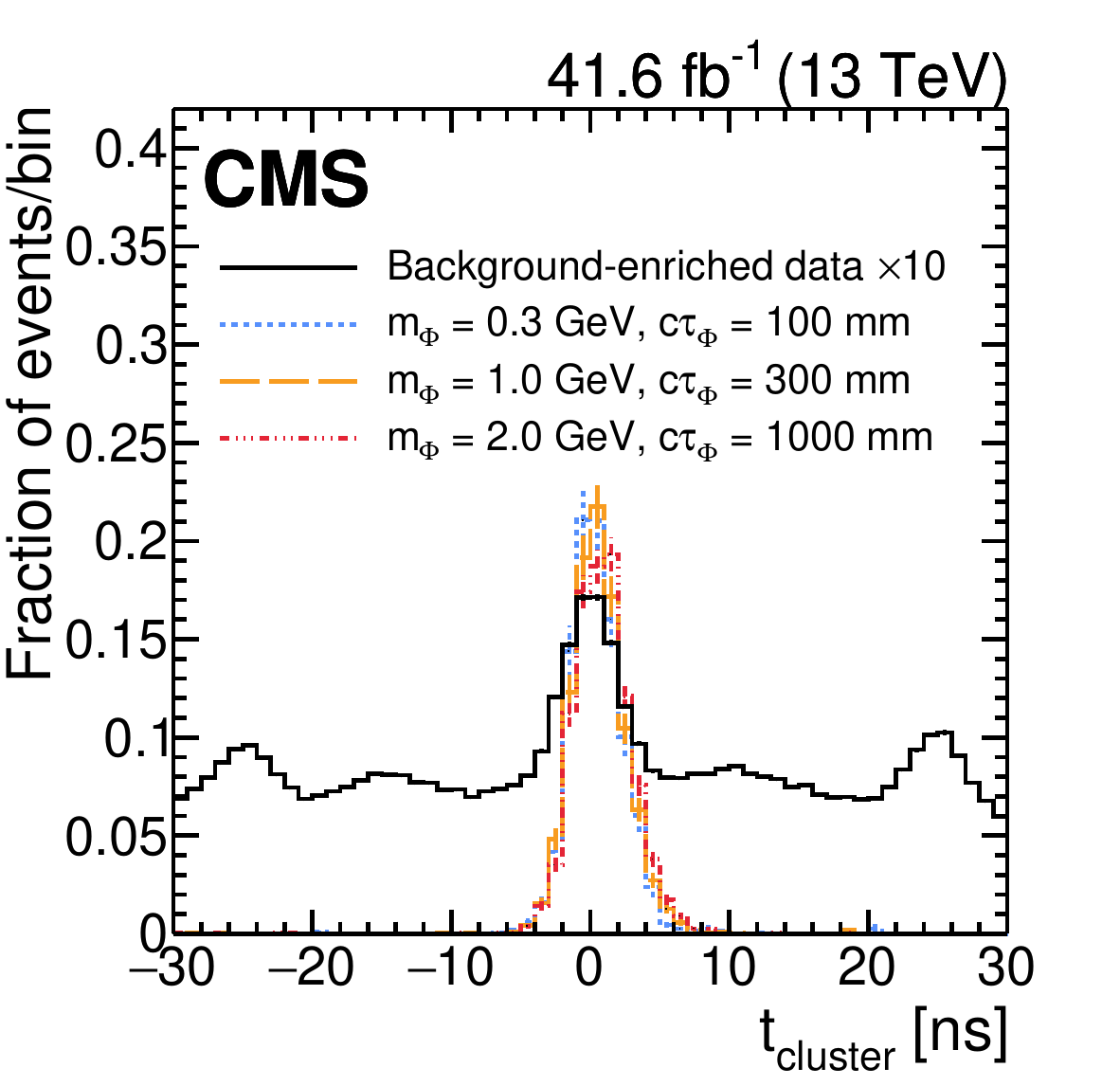}
  \caption{Distribution of $\tcluster$ for the signal and
    the background-enriched data. The distributions are normalized to unity and overflows are not included. To enhance the visibility of fluctuations away from the peak, the background-enriched data distribution has been scaled by a factor of 10.}
  \label{fig:ClusterTime}
\end{figure}

In order to study the distribution of this variable before signal selections are applied, we use a sample of background-enriched data.
The background-enriched data is the sample of data after applying the event level selections and cluster vetoes described above.
The distribution of $\tcluster$ is shown for both the signal and the background-enriched data in Fig.~\ref{fig:ClusterTime}.
The signal peaks sharply near 0\unit{ns}, whereas the background-enriched data also contains contributions from events occurring in adjacent bunch crossings, known as out-of-time (OOT) pileup.
These contributions appear as bumps in the background-enriched data, most prominently at $\tcluster$ values corresponding to the 25\unit{ns} spacing of bunch crossings.
There is a slight asymmetry in the signal $\tcluster$ distribution due to slow moving LLPs, a phenomenon which is enhanced as the mass of the LLP gets larger.
For a cluster to be considered in-time (IT), the corresponding $\tcluster$ is required to be within the range of $-5 < \tcluster< 12.5\unit{ns}$.
The cluster time spread ($\sigma_t$) is defined as $\sigma_t = \sqrt{\smash[b]{\ifrac{\sum_i(t_i - \bar{t})^2}{N}}}$, where $t_i$ is the time of the $i$th hit recorded in the CSC strip layers, $\bar{t}$ is the average time for all hits matched to a cluster, and $N$ is the total number of hits.
A restriction of $\sigma_t< 20$\unit{ns} is applied to avoid clusters which contain hits from multiple bunch crossings. 
The \emph{IT sample} is the collection of events passing all of the above vetoes as well as the cluster time requirements, while the \emph{OOT sample} consists of those passing all of the above vetoes but not the cluster time requirement.
After applying the cluster level selections above, we attain signal efficiency of approximately 10 to 15\% depending on the signal model and a background rejection factor of nearly 2000. 
\begin{figure}[!htb]
  \centering
  \includegraphics[width=0.45\textwidth]{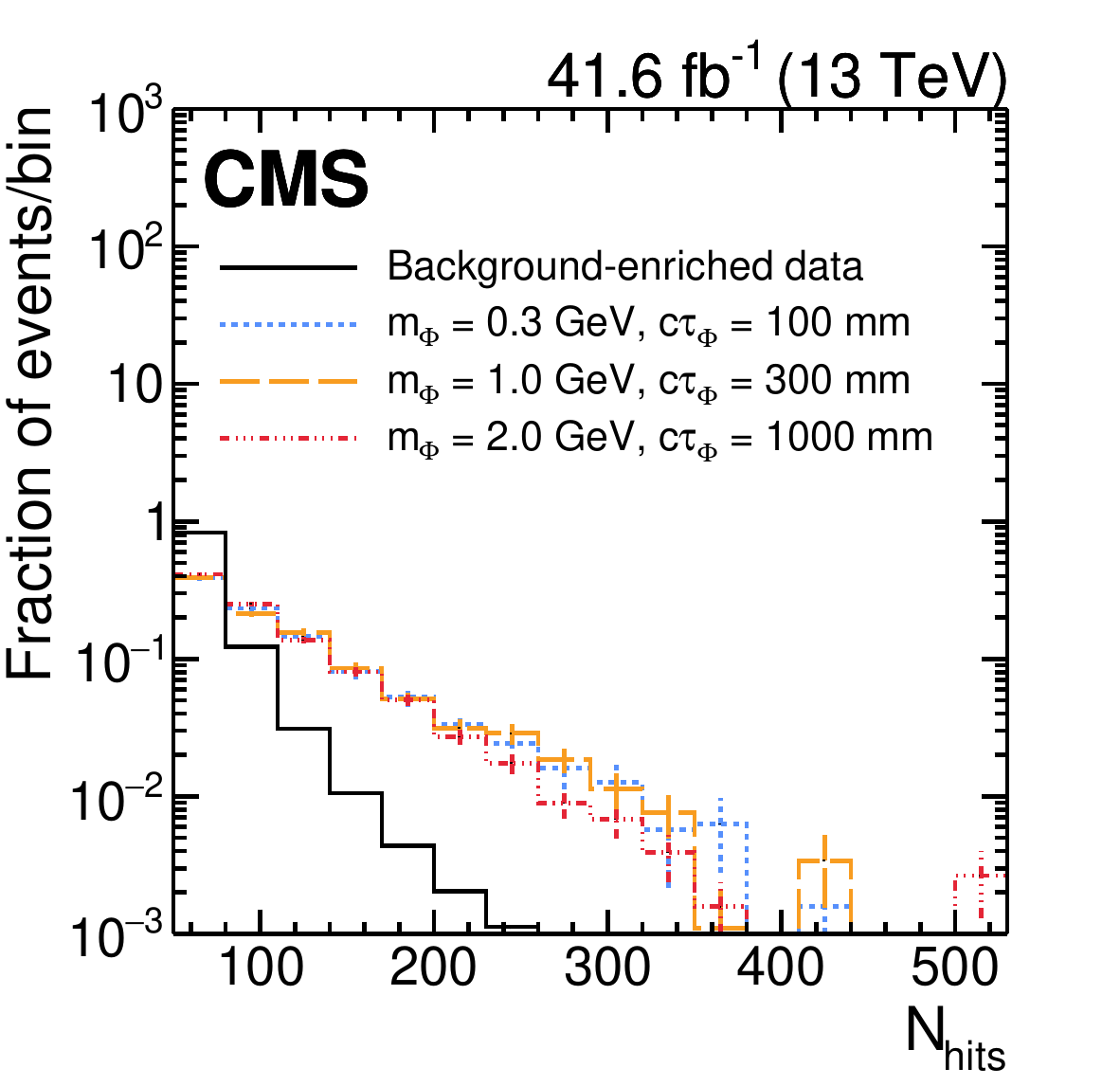}
  \includegraphics[width=0.45\textwidth]{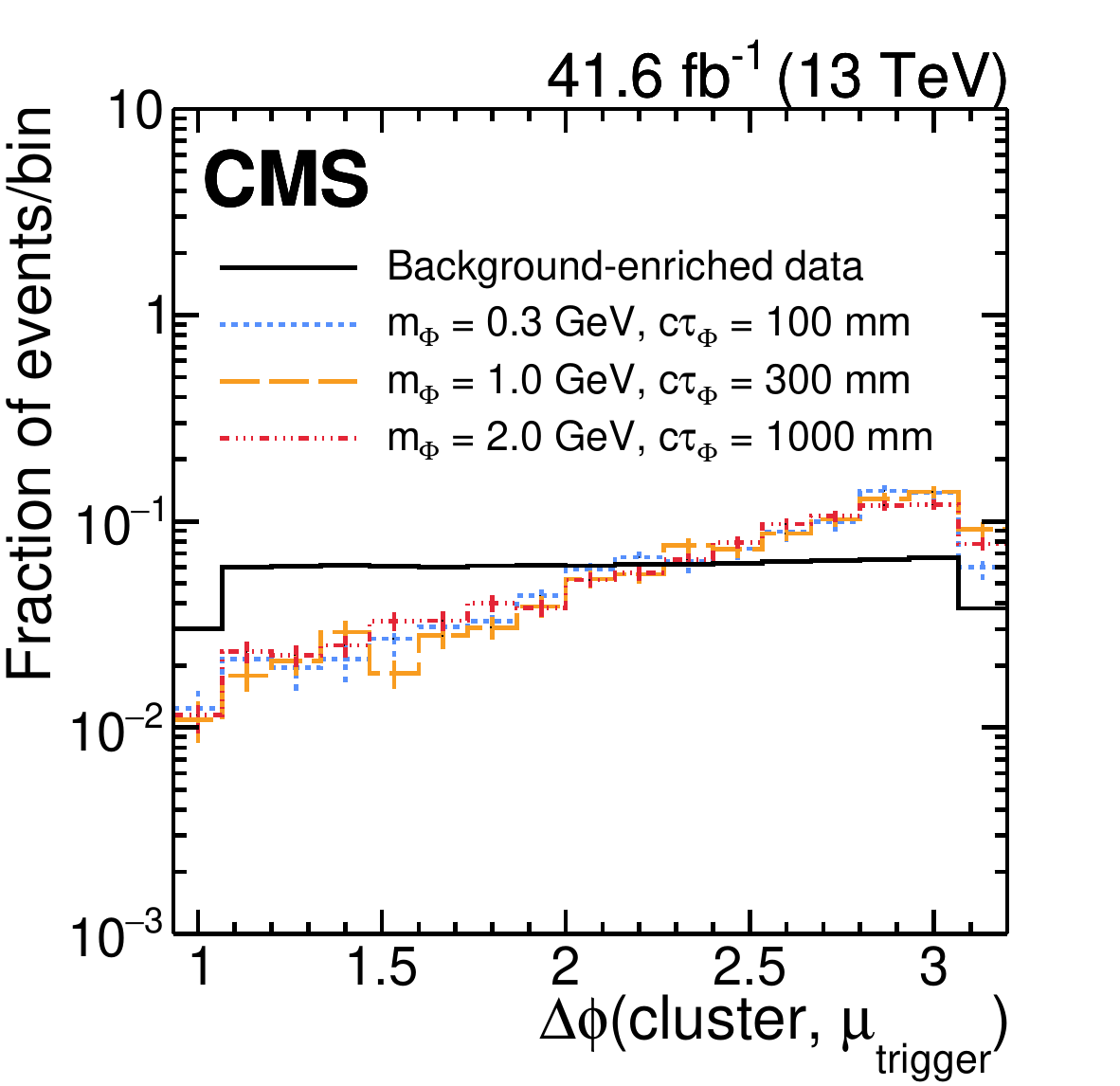}
  \caption{Distributions of the CSC cluster size \Nhits (\cmsLeft) and \vDPhi (\cmsRight), shown
    for various signal samples and the background-enriched data.
    The distributions are normalized to unity and overflows are not included.}
  \label{fig:vars}
\end{figure}

We construct two observables to help distinguish clusters consistent with the signal versus clusters consistent with background processes.
The first of these variables is the cluster size, the number of reconstructed hits comprising the cluster, denoted \Nhits.
The distributions of \Nhits for the background-enriched data and various signal hypotheses are shown in Fig.~\ref{fig:vars} (\cmsLeft).
The second observable measures the location of the cluster with respect to the trigger muon, that is, the difference in the $\phi$ coordinates of the cluster and the muon, \vDPhi.
The distributions of \vDPhi for the background-enriched data and various signal samples are shown in Fig.~\ref{fig:vars} (\cmsRight), where a lower bound on the \vDPhi axis at 1.0 has been applied to reduce the population of clusters that overlap with muons.
Selections on the observables \Nhits and \vDPhi are made in order to distinguish signal from background events.
Signal events generally have a broad \Nhits distribution, extending to relatively large values, while SM background processes exhibit a sharply falling distribution.
For \vDPhi, the signal exhibits a peak at $\pi$, whereas the background distribution is more uniform.
The signal is expected to be produced opposite the triggering muon while the background may be produced anywhere with respect to the triggering muon.

\section{Background estimation}
\label{sec:bkg}

The total background is estimated in two steps. 
First, we estimate the expected contribution of the jet-induced background by measuring the rate at which a jet is misidentified as a signal cluster and applying that misidentification rate to the full \PB parking data set. 
Then, data control samples are used to estimate the size of the uncorrelated background after applying the selections described in the previous Section.
To predict the contribution from the uncorrelated background, we use the ``ABCD'' method, which requires two variables that discriminate between signal and background and are independent of one another for the background~\cite{Kasieczka:2020pil}.
The yields used when making the ABCD prediction reflect the number of events after taking into account the number of jets misidentified as signal clusters.

\begin{figure}[!htb]
  \centering
  \includegraphics[width=\cmsFigWidth]{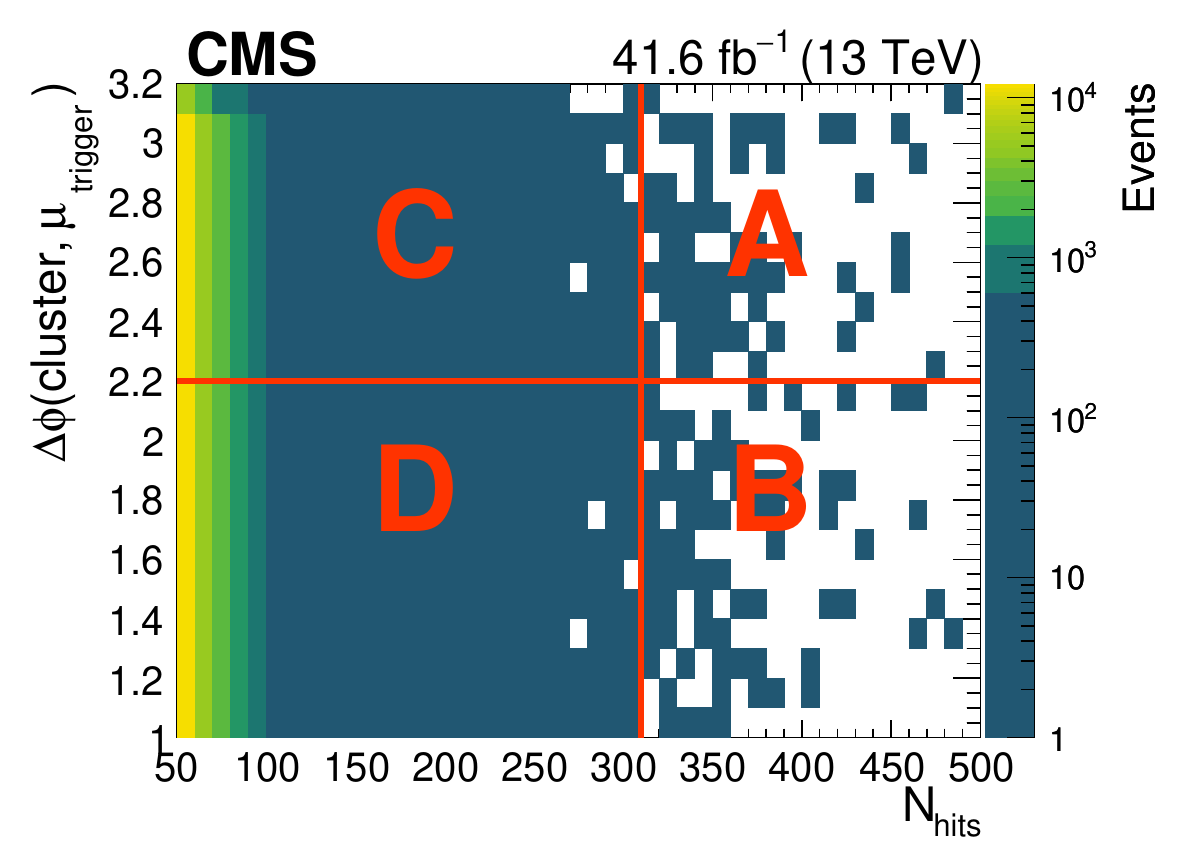}
  \caption{The 2-dimensional distribution of the \Nhits versus \vDPhi for data in the OOT region.}
  \label{fig:Nrechits_vs_dphi}
\end{figure}

The ABCD plane is constructed using \Nhits and \vDPhi as the two observables and is shown in Fig.~\ref{fig:Nrechits_vs_dphi}.
The SR is defined as the upper right quadrant where $\Nhits > 310$ and $\vDPhi > 2.2$.
To determine the definitions of the A (signal), B, C, and D regions, we optimized the signal-to-background ratio in the SR.
The expected number of background events in the SR $N_\mathrm{A}$ can be obtained as $N_\mathrm{A} = N_\mathrm{B} N_\mathrm{C}/N_\mathrm{D}$, where $N_\mathrm{B}$, $N_\mathrm{C}$, and $N_\mathrm{D}$ are the corresponding expected background yields in the B, C, and D regions, respectively.
Further, any potential signal contributions in regions B, C, and D are taken into account through a simultaneous fit of all four regions, where the signal yields in each region are derived independently from simulation but scaled by a common signal strength parameter.

In contrast to the uncorrelated background, the jet-induced background is expected to be correlated in \Nhits and \vDPhi.
These jets mostly arise from SM decays of the \PQb hadrons, which are expected to be produced antialigned with the trigger muon.
The size of the correlated contribution to the total background depends on the proportion of jets that create a shower in the muon system, which is estimated using events enriched in SM jets entering the muon system, provided by a sample of $\PW$+jets.
The background is then subtracted from the yields in regions B, C, and D.

The $\PW$+jets sample is derived from additional data sets collected in 2016--2018 using triggers requiring a single muon, a single electron, or a single photon.
We select events containing jets with $\pt > 20\GeV$ and $\ptmiss > 30\GeV$.
Each event is required to contain at least one isolated lepton (electron or muon) satisfying tight identification criteria.
Jets matched to the isolated, identified lepton are vetoed.
The efficiency with which a jet matches an MDS cluster within $\DR(\text{jet}, \text{cluster}) < 0.4$ is then measured.
Clusters are required to pass the same vetoes and selections as in the SR.
The misidentification rate is measured as a function of jet \pt and minimum cluster size, and then applied to the population of jets in the \PB parking data set.
It was found that, on average, between approximately 4.7 and 100 million jets are needed to produce a single misidentified signal-like cluster, depending on the jet $\pt$.
The measured jet-induced background caused by SM jets producing clusters in the muon system was found to be $156^{+31}_{-22}\stat$.
The listed uncertainty reflects the propagation of statistical uncertainties from both the finite $\PW$+jets sample used to measure the misidentification rate and the finite \PB parking dataset to which this rate is applied for the background estimate.

To validate the misidentification rate estimation procedure for the jet-induced background, we construct two nonindependent validation samples in \Nhits and \vDPhi space and confirm that the ABCD prediction holds when accounting for the additional background source.
We perform this validation in a data region orthogonal to the signal enhanced bin, using regions in the \Nhits, \vDPhi space that fail the SR selection on \vDPhi \ie, the validation regions are subsets of regions B and D as shown in Fig.~\ref{fig:Nrechits_vs_dphi}.
We define two different validation samples that test the accuracy of the jet-induced background prediction in two ranges of extrapolation in the \vDPhi distribution. 

{
\tolerance=800
For validation sample 1, only events in the band $1.5 < \vDPhi < 2.2$ are considered.
Then, the thresholds $\vDPhi = 1.9$ and $\Nhits = 310$ are used to define the boundaries between alternative ABCD regions.
With validation sample 1, we test the jet-induced background prediction extrapolating from the 1.5--1.9 to the 1.9--2.2 regions in \vDPhi. 
Similarly, for validation sample 2, only events in the band $1.0 < \vDPhi < 2.2$ are considered, and $\vDPhi = 1.5$ and $\Nhits = 310$ define the boundaries between alternative ABCD regions.
With validation sample 2, we test the jet-induced background prediction extrapolating from the 1.0--1.5 to the 1.5--2.2 regions in \vDPhi.
\par
}

The jet-induced background (based on the misidentification rate method) and uncorrelated background (based on the ABCD method) are estimated in region A of each validation sample.
These estimates are combined to produce the total background estimate, and compared with the observed number of events in data.
In validation sample 1, $20^{+5}_{-4}$ ($14^{+3}_{-2}$) total (jet-induced) background events are expected and 18 events are observed; while in validation sample 2, $36^{+10}_{-9}$ ($27^{+6}_{-4}$) background events are expected and 40 events are observed.

To validate the performance of the ABCD method used to estimate the size of the uncorrelated background, an OOT sample is defined that targets events arising from pileup interactions from neighboring $\Pp\Pp$ bunch crossings.
The OOT validation sample is constructed by inverting the selection on the $\tcluster$ variable, while retaining the same selections on all other variables.
The jet-induced background is not present in the OOT validation sample.
The ABCD prediction is carried out using the OOT validation sample in the same way as the IT sample.
When performing the validation, we predict $94.6 \pm 8.6$ background events in the OOT sample, region A and observe 95 events.
Figure~\ref{fig:Nrechits_vs_dphi} displays the 2-dimensional distribution of \Nhits and \vDPhi for data in the OOT validation sample, illustrating their independence for background events.

\section{Systematic uncertainties}
\label{sec:unc}

Both the jet-induced and the uncorrelated background estimates have statistical uncertainties based on the measured event yields used in both the misidentification rate and ABCD methods.
Because of the validation presented in the previous Section, no additional systematic uncertainty in the background estimation method is assessed.

A summary of the systematic uncertainties affecting the signal CSC clusters and their sources can be found in Table~\ref{tab:SignalSystematicsCSC}.
The signal yield is affected by the uncertainty in the integrated luminosity, estimated as 2.5\% by a dedicated analysis~\cite{CMS-PAS-LUM-18-002}.
All uncertainties listed in this Section are expressed as percentages relative to the signal efficiency.
The signal yield is also affected by the uncertainty in the cross section.
The cross section for \PBpm production is experimentally measured by the CMS experiment Ref.~\cite{CMS:2024ita} and the inclusive cross section used in this analysis is extrapolated from that measurement based on the known fragmentation
values~\cite{ParticleDataGroup:2022pth}.
We propagated a 15\% uncertainty from that measurement as a systematic uncertainty for our signal normalization. 
When requiring the presence of a muon, a correction factor, referred to as the muon ID scale factor, is applied to account for the differences between the simulated and actual detector response when reconstructing muons. 
This factor affects the signal efficiency and its uncertainty is propagated to the signal yield (0.6\%). 
Finally, the systematic uncertainty from the limited size of the simulated signal sample is in the range 13--26\%.
\begin{table}[!tbph]
  \centering
  \topcaption{Summary of the systematic uncertainties in the signal yield expectations in
    the A, B, C, and D regions for CSC clusters.}
  \label{tab:SignalSystematicsCSC}
  \begin{scotch}{lcccc}
    Source                          & Uncertainty [\%] \\
    \hline          
    Integrated luminosity           & 2.5              \\
    Cross section                   & 15               \\
    Muon ID scale factor            & 0.6              \\
    Simulated signal sample size    & 13--26            \\
    ME11/12 veto                    & 2.2              \\
    RB1 veto                        & 2.6              \\
    RE1 veto                        & 0.7              \\
    MB1 segment veto                & 1.1              \\
    Muon bremsstrahlung veto        & 0.5              \\
    Cluster size correction         & 2.8              \\
    Cluster time                    & 0.3              \\
    Cluster time spread             & 7.8              \\[\cmsTabSkip]
    Total                           & 22.0--31.5       \\
  \end{scotch}
\end{table}

To assess the systematic uncertainties from potential mismodeling in the simulated signal efficiency of the various cluster-level vetoes, a study was performed using both simulated and recorded samples of \Zuu events, where one of the muons underwent bremsstrahlung radiation and produced a cluster in the muon system.
The recorded \Zuu events were selected on-line with a trigger requiring at least one muon with $\pt > 27\GeV$.
Subsequently, muons were required offline to have $\pt > 50\GeV$ for both the simulated and recorded samples.
Each event for both simulated and recorded sample was required to contain a pair of oppositely charged muons with the invariant mass greater than $120\GeV$; this requirement on mass was chosen to enhance the rate of muon bremsstrahlung.
The tag and probe method~\cite{CMS:2011aa} was then used to create a sample of clusters similar to the ones used in the analysis.
The tag (probe) muons were required to pass the tight (loose) muon identification requirements in both the simulated and recorded samples.
The clusters in both simulation and data were required to have a minimum of 50 hits, to match the minimum size requirement used when constructing clusters in the CSCs.
Probe muons were required to match to an MDS cluster with $\DR < 0.4$.
Additionally, in the case of the simulated sample, the probe muon was required to match to a generated muon with $\DR < 0.4$. 
Each veto was applied in the \Zuu data sample and the \Zuu simulated sample.
The ratio of the number of events that passed the tag and probe selections where the probe muon matched an MDS cluster to the number of events passing the tag and probe selections alone defined the cluster efficiency.
The relative difference of the cluster efficiency between data and simulation after applying the various vetoes to MDS clusters was taken as the corresponding systematic uncertainty.
For the ME11/12, RB1, RE1, MB1 segment, and muon bremsstrahlung vetoes, we measured relative differences of 2.2, 2.6, 0.7, 1.1, and 0.5\%, respectively. 
The cluster size distribution falls exponentially and exhibits a shape disagreement between simulation and data, with the distribution for data falling off more rapidly. 
To account for this, a correction factor was applied that shifted the distribution in simulation to the left, improving the agreement in the tail region, where signal is expected to be found.
This corrective factor affected the signal efficiency and its uncertainty was propagated to the signal yield (2.8\%).
Finally, the uncertainties in the cluster time selection and the requirement on the cluster time spread were estimated using the same procedure as for the cluster-level vetoes and were measured to be 0.3 and 7.8\% respectively. 
The uncertainties described in this paragraph have been measured independently in each of the regions defined by the ABCD method used in the previous Section. 
Ultimately, as expected, the size of the uncertainty does not vary much between the different ABCD regions so a representative value is chosen and applied to the signal approximation in all regions for each systematic uncertainty.

\section{Results}
\label{sec:results}
The number of observed events, estimated uncorrelated background, and the estimated jet-induced background for regions B, C, D, and A (SR) are listed in Table~\ref{tab:yields}, along with their corresponding statistical uncertainties.

\newlength{\colwidth}
\setlength{\colwidth}{2.3cm}
\begin{table*}[ht!]
  \centering
  \caption{Summary of the observed event yields and estimated uncorrelated and jet-induced backgrounds in the A (SR), B, C, and D regions.
    The corresponding statistical uncertainties in the estimates are shown.}
  \label{tab:yields}
\begin{scotch}{lllll}
                          & \makebox[\colwidth][c]{B} 
                          & \makebox[\colwidth][c]{C} 
                          & \makebox[\colwidth][c]{D} 
                          & \makebox[\colwidth][c]{A (SR)} \\[\cmsTabSkip]
\hline\noalign{\vskip\cmsTabSkip}
Jet-induced background    & \makebox[\colwidth][c]{$41^{+10}_{-7}$} 
                          & \makebox[\colwidth][c]{$16130^{+240}_{-230}$} 
                          & \makebox[\colwidth][c]{$5701^{+93}_{-91}$} 
                          & \makebox[\colwidth][c]{$156^{+31}_{-22}$}       \\[\cmsTabSkip]
Uncorrelated background   & \makebox[\colwidth][c]{$19 \pm 10$}
                          & \makebox[\colwidth][c]{$42600 \pm 1200$}
                          & \makebox[\colwidth][c]{$51340 \pm 490$}
                          & \makebox[\colwidth][c]{$16 \pm 8$}              \\[\cmsTabSkip]
Total background          & \makebox[\colwidth][c]{$60^{+14}_{-12}$}
                          & \makebox[\colwidth][c]{$58800 \pm 1300$}
                          & \makebox[\colwidth][c]{$57000 \pm 500$}
                          & \makebox[\colwidth][c]{$172^{+32}_{-23}$}       \\[\cmsTabSkip]
Observed events           & \makebox[\colwidth][c]{60}
                          & \makebox[\colwidth][c]{58760}
                          & \makebox[\colwidth][c]{57043}
                          & \makebox[\colwidth][c]{181}                    \\
\end{scotch}
\end{table*}
\begin{figure}[!htb]
  \centering
  \includegraphics[width=\cmsFigWidth]{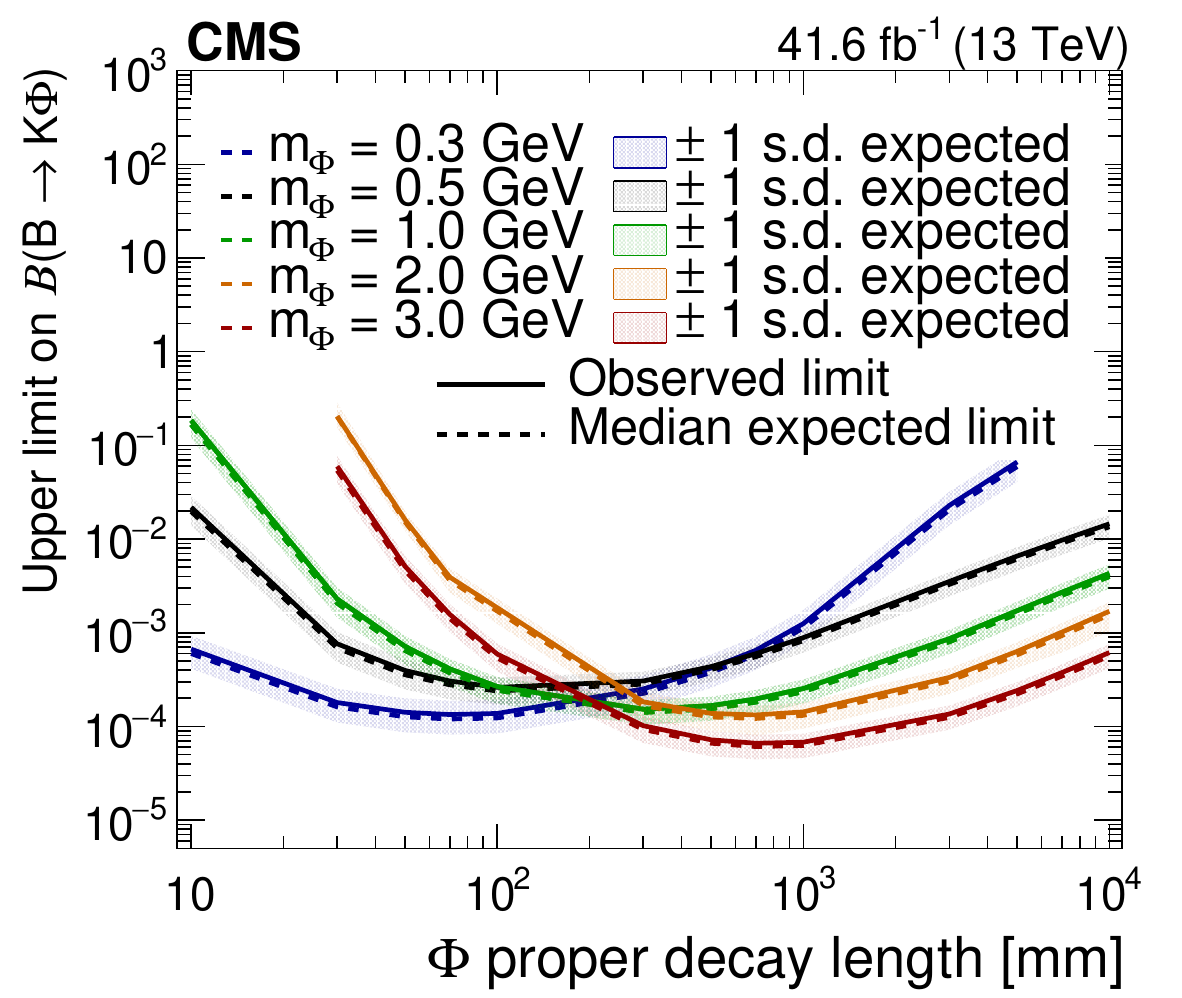}
  \caption{The 95\% \CL exclusion limits on \BF,
    where the scalar LLP, \PGF, decays to a pair of hadrons.
    Limits are presented for scalar
    masses $\mP = 0.3$, 0.5, 1.0, 2.0, and 3.0\GeV as a function of the \PGF mean proper decay length.}
  \label{fig:expLim}
\end{figure}

We observe no significant deviations from the standard model expectation, and we proceed to set 95\% confidence level (\CL) upper limits on the branching fraction for the decay of \PQb hadrons to an LLP, \BF.
Limits are expressed in terms of the inclusive branching fraction \BF, and thus do not depend the decay final state of the LLP. 
This approach separates the production process from the subsequent decay, allowing the results to be reinterpreted in other models by applying the appropriate model-dependent branching fraction.
We follow the LHC \CLs criterion~\cite{CLS2,CLS1} by using the profile likelihood ratio test statistic and the asymptotic formula~\cite{Asymptotic} to evaluate the 95\% \CL observed and expected upper limits on \PGF production and decay.
Results are determined using the CMS statistical analysis tool \textsc{Combine}~\cite{CMS:2024onh}, which is based on the \textsc{RooFit}~\cite{Verkerke:2003ir} and \textsc{RooStats}~\cite{Moneta:2010pm} frameworks.
\begin{figure}[!htb]
  \centering
  \includegraphics[width=\cmsFigWidth]{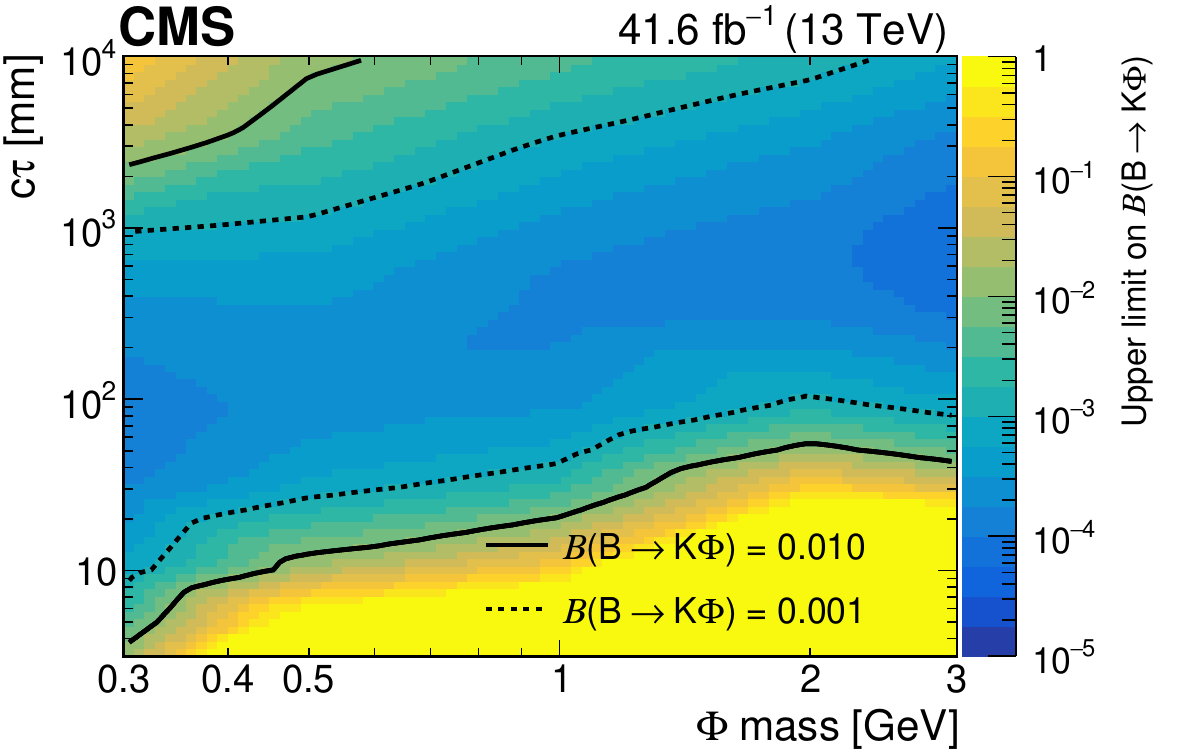}
  \caption{The 95\% \CL exclusion limits on \BF,
    where the scalar LLP, \PGF, decays to a pair of hadrons for different masses and mean proper decay lengths.}
  \label{fig:expLim2D}
\end{figure}

Figure~\ref{fig:expLim} presents the 95\% \CL expected and observed upper limits on \BF for signal models with $\mP = 0.3$, 0.5, 1.0, 2.0, and 3.0\GeV, and where the LLP \PGF decays to a pair of hadrons.
The limits are shown as a function of the \PGF mean proper decay length \ct.
Over most of the range, the upper limit set on \BF is of order $10^{-4}$.
The U-shaped form of the limit plots is caused by a variation in the acceptance efficiency, which is highest when the decay occurs within the muon system, and decreases for lower and higher values of \ct.
When the value of \ct is low (${<}10\unit{cm}$), most of the LLP decays occur before entering the muon system and are thus less likely to produce a large cluster.
Similarly, when \ct is large (${>}10^2\unit{cm}$), a larger fraction of LLPs leave the detector without creating a full cluster.
The limits for the mass range from 0.3 to 3.0\GeV are interpolated based on the calculated values at the masses of 0.3, 0.5, 1.0, 2.0, and 3.0\GeV, and are presented in Fig.~\ref{fig:expLim2D}. 
The sensitivity of the search is dominated by \PGF decays producing hadronic showers.
The contribution to the expected signal yield from \PGF decays resulting in electromagnetic showers is minimal because of the substantially lower energy deposit compared to that from decays producing hadronic showers.

\section{Summary}
\label{sec:summary}

A search for a long-lived particle (LLP) \PGF produced from the decay of a bottom hadron and decaying to a pair of hadrons in the CMS endcap muon system has been carried out.
The search uses the \PB parking data set of proton-proton collisions at $\sqrt{s}=13\TeV$ collected by the CMS experiment in 2018 and corresponding to an integrated luminosity of 41.6\fbinv.
The data set was collected using a set of high-rate triggers that require the presence of a single displaced muon.
A strategy based on control samples in data has been developed to evaluate separately the uncorrelated background originating from prompt standard model particles that produce a shower in the muon system and the jet-induced background from long-lived standard model particles decaying to jets.
The branching fraction \BF is constrained for five low-mass LLP signal hypotheses.
The best sensitivity for measuring a limit on \BF for these models is observed in the range of 1--100\unit{cm} for the \PGF mean proper decay length \ct.
The most sensitive \ct value monotonically increases with LLP mass.
However, the best sensitivity is similar across all models, constraining
\BF to be below $10^{-4}$ at 95\% confidence level depending on the value of 
\ct for each of the signal models studied.
Searches for scalar LLPs produced in the decay of bottom hadrons have been performed by the CHARM, Belle, BaBar, and LHCb Collaborations, with the LHCb search in the dimuon final state providing the most stringent overall bounds.
Complementary to these efforts, the search in this paper targets hadronic decays of the LLP and yields the most stringent limits to date on the decay of a long-lived scalar particle to a pair of hadrons when the LLP is produced by a bottom hadron, particularly for long decay lengths.
\begin{acknowledgments}

We congratulate our colleagues in the CERN accelerator departments for the excellent performance of the LHC and thank the technical and administrative staffs at CERN and at other CMS institutes for their contributions to the success of the CMS effort. In addition, we gratefully acknowledge the computing centers and personnel of the Worldwide LHC Computing Grid and other centers for delivering so effectively the computing infrastructure essential to our analyses. Finally, we acknowledge the enduring support for the construction and operation of the LHC, the CMS detector, and the supporting computing infrastructure provided by the following funding agencies: SC (Armenia), BMBWF and FWF (Austria); FNRS and FWO (Belgium); CNPq, CAPES, FAPERJ, FAPERGS, and FAPESP (Brazil); MES and BNSF (Bulgaria); CERN; CAS, MoST, and NSFC (China); MINCIENCIAS (Colombia); MSES and CSF (Croatia); RIF (Cyprus); SENESCYT (Ecuador); ERC PRG, TARISTU24-TK10 and MoER TK202 (Estonia); Academy of Finland, MEC, and HIP (Finland); CEA and CNRS/IN2P3 (France); SRNSF (Georgia); BMBF, DFG, and HGF (Germany); GSRI (Greece); NKFIH (Hungary); DAE and DST (India); IPM (Iran); SFI (Ireland); INFN (Italy); MSIT and NRF (Republic of Korea); MES (Latvia); LMTLT (Lithuania); MOE and UM (Malaysia); BUAP, CINVESTAV, CONACYT, LNS, SEP, and UASLP-FAI (Mexico); MOS (Montenegro); MBIE (New Zealand); PAEC (Pakistan); MES and NSC (Poland); FCT (Portugal);  MESTD (Serbia); MICIU/AEI and PCTI (Spain); MOSTR (Sri Lanka); Swiss Funding Agencies (Switzerland); MST (Taipei); MHESI and NSTDA (Thailand); TUBITAK and TENMAK (T\"{u}rkiye); NASU (Ukraine); STFC (United Kingdom); DOE and NSF (USA).

\hyphenation{Rachada-pisek} Individuals have received support from the Marie-Curie program and the European Research Council and Horizon 2020 Grant, contract Nos.\ 675440, 724704, 752730, 758316, 765710, 824093, 101115353, 101002207, 101001205, and COST Action CA16108 (European Union); the Leventis Foundation; the Alfred P.\ Sloan Foundation; the Alexander von Humboldt Foundation; the Science Committee, project no. 22rl-037 (Armenia); the Fonds pour la Formation \`a la Recherche dans l'Industrie et dans l'Agriculture (FRIA-Belgium); the Beijing Municipal Science \& Technology Commission, No. Z191100007219010, the Fundamental Research Funds for the Central Universities, the Ministry of Science and Technology of China under Grant No. 2023YFA1605804, and the Natural Science Foundation of China under Grant No. 12061141002 (China); the Ministry of Education, Youth and Sports (MEYS) of the Czech Republic; the Shota Rustaveli National Science Foundation, grant FR-22-985 (Georgia); the Deutsche Forschungsgemeinschaft (DFG), among others, under Germany's Excellence Strategy -- EXC 2121 ``Quantum Universe" -- 390833306, and under project number 400140256 - GRK2497; the Hellenic Foundation for Research and Innovation (HFRI), Project Number 2288 (Greece); the Hungarian Academy of Sciences, the New National Excellence Program - \'UNKP, the NKFIH research grants K 131991, K 133046, K 138136, K 143460, K 143477, K 146913, K 146914, K 147048, 2020-2.2.1-ED-2021-00181, TKP2021-NKTA-64, and 2021-4.1.2-NEMZ\_KI-2024-00036 (Hungary); the Council of Science and Industrial Research, India; ICSC -- National Research Center for High Performance Computing, Big Data and Quantum Computing, FAIR -- Future Artificial Intelligence Research, and CUP I53D23001070006 (Mission 4 Component 1), funded by the NextGenerationEU program (Italy); the Latvian Council of Science; the Ministry of Education and Science, project no. 2022/WK/14, and the National Science Center, contracts Opus 2021/41/B/ST2/01369 and 2021/43/B/ST2/01552 (Poland); the Funda\c{c}\~ao para a Ci\^encia e a Tecnologia, grant CEECIND/01334/2018 (Portugal); MICIU/AEI/10.13039/501100011033, ERDF/EU, "European Union NextGenerationEU/PRTR", and Programa Severo Ochoa del Principado de Asturias (Spain); the Chulalongkorn Academic into Its 2nd Century Project Advancement Project, and the National Science, Research and Innovation Fund via the Program Management Unit for Human Resources \& Institutional Development, Research and Innovation, grant B39G680009 (Thailand); the Kavli Foundation; the Nvidia Corporation; the SuperMicro Corporation; the Welch Foundation, contract C-1845; and the Weston Havens Foundation (USA).
\end{acknowledgments}

\bibliography{auto_generated}
\cleardoublepage \appendix\section{The CMS Collaboration \label{app:collab}}\begin{sloppypar}\hyphenpenalty=5000\widowpenalty=500\clubpenalty=5000\input{EXO-24-004-public-authorlist.tex}\end{sloppypar}
\end{document}

%% file: EXO-24-004-public-authorlist.tex
\cmsinstitute{Yerevan Physics Institute, Yerevan, Armenia}
{\tolerance=6000
A.~Hayrapetyan, V.~Makarenko\cmsorcid{0000-0002-8406-8605}, A.~Tumasyan\cmsAuthorMark{1}\cmsorcid{0009-0000-0684-6742}
\par}
\cmsinstitute{Institut f\"{u}r Hochenergiephysik, Vienna, Austria}
{\tolerance=6000
W.~Adam\cmsorcid{0000-0001-9099-4341}, J.W.~Andrejkovic, L.~Benato\cmsorcid{0000-0001-5135-7489}, T.~Bergauer\cmsorcid{0000-0002-5786-0293}, M.~Dragicevic\cmsorcid{0000-0003-1967-6783}, C.~Giordano, P.S.~Hussain\cmsorcid{0000-0002-4825-5278}, M.~Jeitler\cmsAuthorMark{2}\cmsorcid{0000-0002-5141-9560}, N.~Krammer\cmsorcid{0000-0002-0548-0985}, A.~Li\cmsorcid{0000-0002-4547-116X}, D.~Liko\cmsorcid{0000-0002-3380-473X}, M.~Matthewman, I.~Mikulec\cmsorcid{0000-0003-0385-2746}, J.~Schieck\cmsAuthorMark{2}\cmsorcid{0000-0002-1058-8093}, R.~Sch\"{o}fbeck\cmsAuthorMark{2}\cmsorcid{0000-0002-2332-8784}, D.~Schwarz\cmsorcid{0000-0002-3821-7331}, M.~Shooshtari, M.~Sonawane\cmsorcid{0000-0003-0510-7010}, W.~Waltenberger\cmsorcid{0000-0002-6215-7228}, C.-E.~Wulz\cmsAuthorMark{2}\cmsorcid{0000-0001-9226-5812}
% ...
\par}
\vskip\cmsinstskip
\dag:~Deceased\\
$^{1}$Also at Yerevan State University, Yerevan, Armenia\\
$^{2}$Also at TU Wien, Vienna, Austria\\
% ...
